# Shorter Communication
# *q*-Exponential Structure of Arbitrary-Order Reaction Kinetics

Robert K. Niven[1]

[1] School of Aerospace, Civil and Mechanical Engineering, The University of New South Wales at ADFA, Northcott Drive, Canberra, ACT, 2600, Australia.
Email: r.niven@adfa.edu.au

20 May 2005

**Abstract**

The rate equation for an arbitrary *m*th order growth or decay reaction can be expressed in terms of the *q*-exponential function $\exp_q f = [1+(1-q)f]^{1/(1-q)}$, with *q* equal to *m*. The analysis suggests that a wide variety of reaction rate (kinetic) processes and models, in chemistry, biology/ecology, unit operations and contaminant transport, are amenable to analysis by Tsallis' non-extensive statistical mechanics. A number of *q*-analogues of common kinetic mathematical models are proposed, and illustrated by several examples.

1. **Introduction**

In recent years, a major advance has taken place in theoretical physics of great importance to engineers, with the development of so-called "non-extensive" statistical mechanics based on the Tsallis (1988) entropy function:

$$S_q = -k_B \sum_{i=1}^{s} p_i^q \ln_q p_i = \frac{k_B}{q-1}\left(1 - \sum_{i=1}^{s} p_i^q\right), \qquad (1)$$

where $S_q$ is the Tsallis entropy, $k_B$ is the Boltzmann (or any other applicable) constant, $p_i$ is the probability of occurrence of the $i$th state, $s$ is the number of possible states, $q \in \mathbb{R}$ is the Tsallis parameter, and $\ln_q f = (1-q)^{-1}(f^{1-q} - 1)$, $f>0$ is the $q$-logarithmic function (Tsallis, 1994). Note $S_q$ is here defined in the units of $k_B$, i.e. J K$^{-1}$ molecule$^{-1}$. In the limit as $q \to 1$, $\ln_q f \to \ln f$ and $S_q$ reduces to the well-known Boltzmann (1877)-Planck (1914) and Shannon (1948) definitions of entropy:

$$S = -k_B \sum_{i=1}^{s} p_i \ln p_i . \qquad (2)$$

Using the Tsallis entropy, a new theory of statistical mechanics (including thermodynamics and information theory) has been built up over the past decade (e.g. Tsallis, 2001; Plastino, 2004) - forming a superset of "traditional" or Maxwell-Boltzmann statistical mechanics - for the analysis of systems which exhibit some form of "correlated" structure between individual entities or elements within the system. Examples include systems with long-range interactions, long-range memory effects and/or multi-fractal (power-law) space-time structures, encompassing a rich assortment of physical, chemical, astronomical, fluid mechanics, engineering and financial systems (Tsallis, 2001; Tsallis *et al.* 2002; Beck, 2000).

The purpose of this communication is to extend the work of several workers (e.g. Tsallis, 2001; Tsallis *et al.*, 1999; Plastino *et al.*, 2000), to delineate another class of systems with $q$-exponential structure, involving rate processes of arbitrary reaction order. Such systems extend over the breadth of physical chemistry and biochemistry; mathematical biology and ecology; chemical, environmental and process engineering; and the migration of constituents (contaminants) within natural or constructed media.

**2.    Theoretical Analysis**

Consider the reaction of a single chemical, biological, radioactive or other species A:

$$A \rightarrow products \quad or \quad reactants \rightarrow A, \tag{3}$$

for which the reaction rate is expressed by the simple arbitrary-order kinetic equation (e.g., Moore, 1972; Levenspiel, 1972; Jordan, 1979; Moore & Pearson, 1981; Atkins, 1982):

$$\frac{d\chi_A}{dt} = k\chi_A^m, \tag{4}$$

where $\chi_A(t) = C_A(t)/C_{Aref}$ is the normalised concentration (more accurately, the normalised activity or fugacity) of species A, such that the activity $C_A(t)$ of species A is expressed relative to some reference activity $C_{Aref}$; $k$ is the rate constant; $t$ is time; and $m$ is the reaction order.  The normalisation of $C_A(t)$ is necessary to avoid expressing $m$ in awkward units; also in general $0 \leq \chi_A(t) \lessgtr 1$ (if $\chi_A(t)$ is a mole or population fraction, then $0 \leq \chi_A(t) \leq 1$).  Eq. (4) with $m=1$ and $k>0$ (or $k<0$) gives the familiar *first order growth* (*decay*) process, common in chemical and biological reactions and always evident in radioactive decay phenomena. Other well-known examples from chemical kinetics include $m=0$ (*zeroth order*) and $m=2$ (*second order*) reactions.   In general $m>0$ and $k>0$ ($k<0$) indicates a growth (decay) relationship of arbitrary $m$th order.

Solution of eq. (4) with the initial condition $\chi_A(0) = \chi_{A0}$ yields (Atkins, 1982; Levenspiel, 1972; Moore & Pearson, 1981; Tsallis *et al.*, 1999):

$$\chi_A(t) = \left[\chi_{A0}^{1-m} + kt(1-m)\right]^{1/(1-m)}. \tag{5}$$

This can be expressed in terms of the *q-exponential function* $\exp_q f = [1+(1-q)f]^{1/(1-q)}$, the inverse of the *q*-logarithm function (Tsallis, 1994):

$$\chi_A(t) = \chi_{A0} \exp_m(\chi_{A0}^{1/(1-m)} kt). \qquad (6)$$

In other words, an arbitrary *m*th order growth or decay process exhibits a *q*-exponential mathematical structure, with *q=m*. If for some reason we had chosen the reference activity $C_{Aref}$ equal to the initial activity $C_{A0}$, whence $\chi_{A0} = 1$, then eq. (6) reduces to:

$$\chi_A(t)\big|_{C_{Aref}=C_{A0}} = \exp_m(kt). \qquad (7)$$

In general, the parameters *k* and $\chi_{A0}$ in eqs. (6)-(7) can be calculated from reaction rate data by the use of various *m-logarithmic* plots of $\ln_m C_A(t)$ against *t*, for different values of *m*.

### 3. Applications

The above rate expressions have obvious implications for the application of Tsallis (power law) statistics to a variety of reaction rate processes of arbitrary order. There are myriad applications in many fields, for example:

- *Chemical kinetics*: *q*-formulations of chemical reaction kinetics, e.g. for the generalised reaction:

$$aA + bB \rightleftharpoons cC + dD, \qquad (8)$$

where A, B, C, D are chemical species and *a*, *b*, *c*, *d* their respective stoichiometric coefficients, we write the rate equation (Moore, 1972; Levenspiel, 1972; Jordan, 1979; Moore & Pearson, 1981; Atkins, 1982):

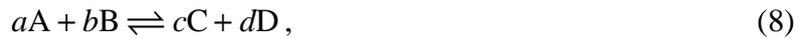

$$\frac{d\xi}{dt} = -\frac{1}{a}\frac{d\chi_A}{dt} = -\frac{1}{b}\frac{d\chi_B}{dt} = \frac{1}{c}\frac{d\chi_C}{dt} = \frac{1}{d}\frac{d\chi_D}{dt} = k\chi_A^\alpha \chi_B^\beta \chi_C^\gamma \chi_D^\delta, \qquad (9)$$

where $\xi$ is a measure of the extent of reaction; $\chi_A, \chi_B, \chi_C, \chi_D$ are the normalised activities of each species and $\alpha, \beta, \gamma, \delta$ are the reaction orders with respect to each species as measured by experiment, in general unequal to unity and not necessarily related to *a*, *b*, *c*, *d* or each other. By stoichiometry, eqs. (8)-(9) give:

$$\frac{d\xi}{dt} = k(\chi_{A0} - a\xi)^\alpha (\chi_{B0} - b\xi)^\beta (\chi_{C0} + c\xi)^\gamma (\chi_{D0} + d\xi)^\delta, \quad (10)$$

where $\chi_{A0}$, $\chi_{B0}$, $\chi_{C0}$, $\chi_{D0}$ are the initial activities of each species. Eqs. (8)-(10) can be further generalised to more complicated multi-reaction chemical systems, for example with opposing, consecutive and/or parallel reactions, as well as multiple or oscillating equilibria (Moore, 1972; Levenspiel, 1972; Jordan, 1979; Moore & Pearson, 1981; Atkins, 1982).

- *Biological and ecological population dynamics:* *q*-variants of the many biological population dynamics models (c.f. Jones & Sleeman, 1983; Krebs, 1985; Banks, 1994; Kapur & Kesevan 1992), for example:

    - simple *q*-exponential growth and decay models, of similar form to eqs. (3)-(4);

    - *q*-logistic models, reflecting some physical limitation to growth, such as (c.f. Jones & Sleeman, 1983; Krebs, 1985; Banks, 1994):

    $$\frac{d\chi_A}{dt} = k_{max}\left(1 - \frac{\chi_A}{K_A}\right)^\kappa \chi_A^\alpha, \quad (11)$$

    where $k_{max}$ is the maximum rate constant, $K_A$ is the normalised carrying capacity (the maximum attainable $\chi_A$) and $\alpha$, $\kappa$ are power law exponents;

    - *q*-Monod models, reflecting limitations to growth due to scarcity of some critical substrate (nutrient or resource), e.g. (c.f. Banks, 1994):

    $$\frac{d\chi_A}{dt} = k_{max}\left(\frac{S}{K_{sA} + S}\right)^\sigma \chi_A^\alpha, \quad (12)$$

    where *S* is the substrate concentration, $K_{sA}$ is the half-saturation coefficient and $\alpha$, $\sigma$ are power law exponents;

    - *q*-interacting species models, including:

(i) *q*-predator-prey models, e.g. for population counts *N* of prey and *P* of predator (c.f. Jones & Sleeman, 1983; Krebs, 1985):

$$\frac{dN}{dt} = k_N N^\nu \left(1 - \frac{N}{K}\right)^\kappa - P^\pi \, F(N,P)$$
$$\frac{dP}{dt} = -k_P P^\pi + N^\nu \, G(N,P)$$
(13)

where $k_N$, $k_P$ are the rate constants for predator and prey; *K* is the carrying capacity of the prey; F(*N*,*P*), G(*N*,*P*) are specified functions of *N* and *P*; and ν, κ and π are power law exponents; and

(ii) *q*-competing species models, e.g. for population counts *M*, *N* (c.f. Jones & Sleeman, 1983; Krebs, 1985):

$$\frac{dM}{dt} = k_M M^\mu \left(1 - \frac{M + aN}{K_M}\right)^\kappa$$
$$\frac{dN}{dt} = k_N N^\nu \left(1 - \frac{N + bM}{K_N}\right)^\lambda$$
(14)

where $k_M$, $k_N$ are rate constants; $K_M$, $K_N$ are carrying capacities; *a*, *b* are model parameters and μ, ν, κ, λ are power law exponents;

- a wide range of other variants of the above, for example with time lag effects, memory effects, age structure or stochastic variables (c.f. Jones & Sleeman, 1983; Krebs, 1985; Banks, 1994).

- *Unit operations:* *q*-formulations of the conservation of mass of each constituent within a control volume in chemical, environmental or process engineering, as given by (Levenspiel, 1972; Mihelcic *et al.* 1999; Felder & Rousseau, 2000):

$$\frac{\partial M_i}{\partial t} = \frac{\partial M_{i,in}}{\partial t} - \frac{\partial M_{i,out}}{\partial t} + \frac{\partial M_{i,react}}{\partial t}, \quad i = 1,...,w$$
(15)

where $\partial M_i / \partial t$ is the net rate of change of mass of constituent *i* (from a total of *w*

constituents) within a control volume; and $\partial M_{i,in}/\partial t$, $\partial M_{i,out}/\partial t$ and $\partial M_{i,react}/\partial t$ are respectively the total rates of flow of mass of $i$ into, out of and production (by reaction) within the control volume. For a constant volume process:

$$\frac{\partial M_i}{\partial t} = V\frac{\partial C_i}{\partial t}, \quad \frac{\partial M_{i,react}}{\partial t} = V\frac{\partial C_{i,react}}{\partial t} \qquad (16)$$

where $V$ is the control volume, and $\partial C_i/\partial t$ and $\partial C_{i,react}/\partial t$ are respectively the net and reactive rates of change of concentration (in mass/volume units) of $i$ in the control volume. For an arbitrary order reaction process, eq. (4) gives (ignoring activity effects):

$$\frac{\partial C_{i,react}}{\partial t} = kC_{ref}\left(\frac{C(t)}{Cref}\right)^m \qquad (17)$$

Eqs. (15)-(17) can be readily applied to typical unit operation problems involving batch, well-mixed or plug flow reactors.

The above equations reduce to their first-order counterparts when the power law exponents approach unity. Some of the above power law models have been presented previously (e.g. Kapur & Kesavan, 1992; Banks, 1994), but apart from a few instances (e.g. Plastino *et al.*, 2000) have not been examined from the perspective of a Tsallisian or "correlated" statistical framework. Furthermore, whilst many of the above models cannot at present be solved analytically, or are of complicated mathematical form, analysis in light of the *q*-exponential structure may reveal superior numerical reduction techniques and/or more concise analytical solutions.

In the above systems, any observed Tsallis-like mathematical structure is not the result of some anomalous (long-range correlated) configuration in geometric space, but rather is brought about by the *reaction mechanism*; i.e. by the structure of reactions within and between individual entities in the system. It is interesting that first order (*m*=1) processes - widely adopted as the default case throughout science and engineering - reflect the "no feedback" scenario; i.e. the probability of reaction of a given entity (individual molecule or organism) is unaffected by the presence of other entities. A non-first order ($m \neq 1$) reaction rate reveals the existence of some interaction mechanism, such that the reaction rate is modified (decelerated or accelerated) by other entities.

## 4. Examples

Several examples will now be considered. Firstly, consider a simple $q$-exponential decay relationship (eqs. (5)-(6)) for a reactive species A, with $C_{A0}=100$ mg L$^{-1}$ (or any other appropriate units), $C_{A,ref}=1$ mg L$^{-1}$ and $k=-220$ min$^{-1}$ (recall $k<0$ indicates a decay relationship). Plots of ln $\chi_A(t)$ against time are illustrated in Figure 1 for various values of $m$. As expected, ln $\chi_A(t)$ for $m=1$ follows a straight line decay relationship with time. For $m<1$ ($m>1$) the reaction proceeds less (more) rapidly, producing a curve of increasing (decreasing) negative gradient with time on a natural logarithm plot.

Now consider a well-mixed tank problem, in which a reactor of constant volume $V=500$ L and steady flow throughput $Q=Q_{in}=Q_{out}=50$ L min$^{-1}$ (hence a mean residence time of $\theta=V/Q=10$ min) initially contains a steady state concentration $C_{A0}=100$ mg L$^{-1}$ of a reactive constituent A, again with a $q$-exponential rate constant of $k=-220$ min$^{-1}$. The incoming concentration is suddenly changed to $C_{in}=0$ at $t=0$, whilst the flow rate $Q$ is kept constant. How will the concentration change with time? We again use a reference concentration of $C_{A,ref}=1$ mg L$^{-1}$, and ignore activity effects. From eqs. (15)-(17), the governing differential equation is:

$$\theta \frac{d\chi_A(t)}{dt} = -\chi_A(t) + k\theta \, [\chi_A(t)]^m \qquad (18)$$

Solution with the initial condition $\chi_A(0) = \chi_{A0}$ yields:

$$\chi_A(t) = \left[ k\theta\left(1 - (e^{-t/\theta})^{(1-m)}\right) + \left(\chi_{A0} e^{-t/\theta}\right)^{(1-m)} \right]^{1/(1-m)}, \tag{19}$$

This is of somewhat similar form to the *q*-exponential function (compare eq. (5)), and clearly belongs to the same mathematical family. Eq. (19) is plotted for the above parameter values and various values of *m* in Figure 2. Once again, $\chi_A(t)$ for *m*=1 exhibits an exponential decay relationship with time. The curves for *m*<1 (*m*>1) follow broadly similar patterns of curvature to those in Figure 1. (A second branch of the *m* = ½ curve is evident for *t*>34.2596 min.; this is mathematically valid but physically spurious.) In this case, however, the decay curves for a reactive species are bounded by the normalised concentration of a non-reactive (conservative) species $\chi_A(t) = \chi_{A0} e^{-t/\theta}$, shown as a dotted line, which is removed solely by dilution.

Now consider a biological species A which exhibits *q*-logistic growth (eq. (11)). With the initial condition $\chi_A(0) = \chi_{A0}$, the (implicit) analytical solution is:

$$t = \frac{\chi_A^{1-m} F\left([\kappa, 1-m], [2-m], \frac{\chi_A}{K_A}\right) - \chi_{A0}^{1-m} F\left([\kappa, 1-m], [2-m], \frac{\chi_{A0}}{K_A}\right)}{k_{\max}(1-m)} \tag{20}$$

where $F([..],[..],..)$ is the generalised hypergeometric function (Waterloo Maple Inc., 1981-2004). For $\kappa = m = 1$, this reduces to the familiar solution:

$$t = \frac{1}{k_{\max}} \ln\left( \frac{\chi_A(K_A - \chi_{A0})}{\chi_{A0}(K_A - \chi_A)} \right) \tag{21}$$

For a specific example with $\chi_{A0}$=1/2500 units, $K_A$=1 units and $k_{\max}$=1 min$^{-1}$, plots of $\chi_A(t)$ obtained using eqs. (20)-(21) are shown in Figures 3a-b - respectively for $\kappa$=1 and *m*=1 - for convenience using a logarithmic time scale. As evident, parameter *m* (Fig. 3a) modifies the steepness of the sigmoidal curve, producing a delayed but more rapid (earlier but less rapid) reaction for *m*>1 (*m*<1). On the other hand, parameter $\kappa$ (Fig. 3b) modifies the rate at which $\chi_A$ approaches its asymptotic limit $K_A$, such that it takes much longer as $\kappa$>>1 (but $\chi_A$ flies off to infinity as $\kappa \to 0$). Both $m \neq 1$ and $\kappa \neq 1$ therefore reflect processes in which the reproduction of species A is modified by the presence of other members of that species, in two different ways.

## 5. Further Discussion

Although not the object of this discussion, it is worth reporting that a *q*-variant of the Arrhenius (1889) expression for the rate constant of a chemical reaction has recently been presented (Andricioaei & Straub, 2001; Lenzi *et al.*, 2001):

$$k_q = A \frac{\exp_q(-E^\ddagger / k_B T)}{\exp_q(-E_0 / k_B T)} \qquad (22)$$

where *A* is a normalising ("pre-*q*-exponential") factor, *T* is absolute temperature, $E^\ddagger$ is the energy of the transition state and $E_0$ is that of the ground state. As $q \to 1$, $k_q$ approaches the familiar Arrhenius (1889) form $k = A\exp(-(E^\ddagger - E_0)/k_B T)$, where $\Delta E = E^\ddagger - E_0$ is the activation energy. It is important that the two different Tsallisian influences on chemical reactions be clearly distinguished: the *q* in eq. (22) reflects a reaction with a Tsallisian (correlated) thermodynamic structure, whilst the power law exponent in the rate expression itself (eqs. (4) or (9)-(14)) reflects a Tsallisian (non-first order) reaction mechanism. Conceivably, there are four possible combinations of these two quite different effects.

The foregoing rate equations can be further incorporated into a generalised nonlinear advection-dispersion-reaction equation (equivalent to a nonlinear Fokker-Planck equation with reaction) of the following type (c.f. Tsallis & Bukman, 1996; Plastino *et al.*, 2000; Lenzi *et al.*, 2001; Plastino, 2001; Frank, 2002):

$$\frac{\partial \chi_i}{\partial t} = -\nabla \cdot (\mathbf{v}\chi_i^\upsilon) + \nabla \cdot (\mathbf{D} \cdot \nabla \chi_i^\delta) + J_i, \quad i = 1,...,w \qquad (23)$$

where $\chi_i = \chi_i(x,y,z,t)$ is the normalised activity of species *i* (entities of type *i*) in cartesian coordinates, from a total set of *w* species; **v** is the velocity (transport) vector; **D** is the dispersion or diffusion coefficient (diffusivity) tensor; $\nabla$ is the grad operator; $\nabla \cdot$ is the div operator; $\upsilon$ and $\delta$ are power law exponents (in general $\upsilon \neq \delta$); and $J_i$ is a generalised non-linear reaction rate equation for species *i*, e.g. of the form given in eqs. (4) or (9)-(14). Eq. (23) provides a *q*-generalised model for the anomalous transport, anomalous (non-Fickian, correlated) dispersion or diffusion, and non-first-order (power-law) growth or decay of any set of physical, chemical and/or biological species. Additional source-sink terms for the creation or destruction (or entry or exit) of entities, if required, can be added to eq. (23). Furthermore, if the advection or dispersion terms for the *i*th species are influenced by the activities of species *j≠i*, time lag effects or other complications, further extensions to eq. (23) are possible.

**Conclusions**

It is shown that the rate equation for an arbitrary *m*th order growth or decay reaction can be expressed in terms of the *q*-exponential function, with *q* equal to *m*. Arbitrary order reactions therefore fall within the framework of Tsallis (correlated) statistical mechanics. The analysis has application in many fields, e.g. physical chemistry and biochemistry; mathematical biology and ecology; chemical, environmental and process engineering; and the migration of constituents (contaminants) within natural or constructed media. Arbitrary order chemical reaction mechanisms are well known; it will be interesting to see if analogous power law kinetic models could be relevant in biological/ecological, process engineering or contaminant transport modelling applications.

**Figure Captions**

Figure 1: Plots of ln $\chi_A(t)$ against time for $m$th order decay of reactive species A, for various values of $m$ (eqs. (5)-(6); for parameter values see text).

Figure 2: Plots of ln $\chi_A(t)$ against time in a well-mixed reactor containing $m$th order reactive species A, for various values of $m$ (eq. (19); for parameter values see text). The dilution of a conservative species is shown by a dashed line.

Figure 3: Plots of $\chi_A(t)$ against ln($t$) for $q$-logistic growth of biological species A (eqs. (20)-(21)), for (a) $\kappa=1$ and various $m$; and (b) $m=1$ and various $\kappa$ (for parameter values see text).

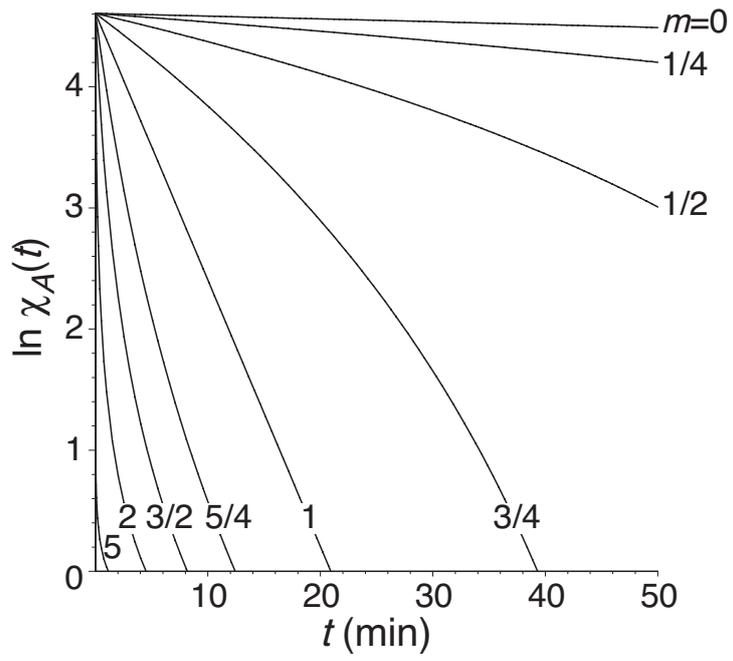

FIGURE 1

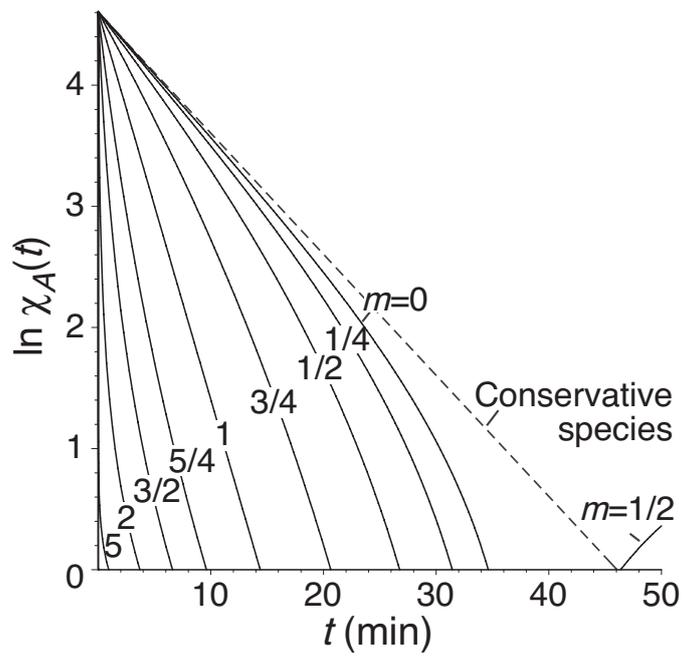

FIGURE 2

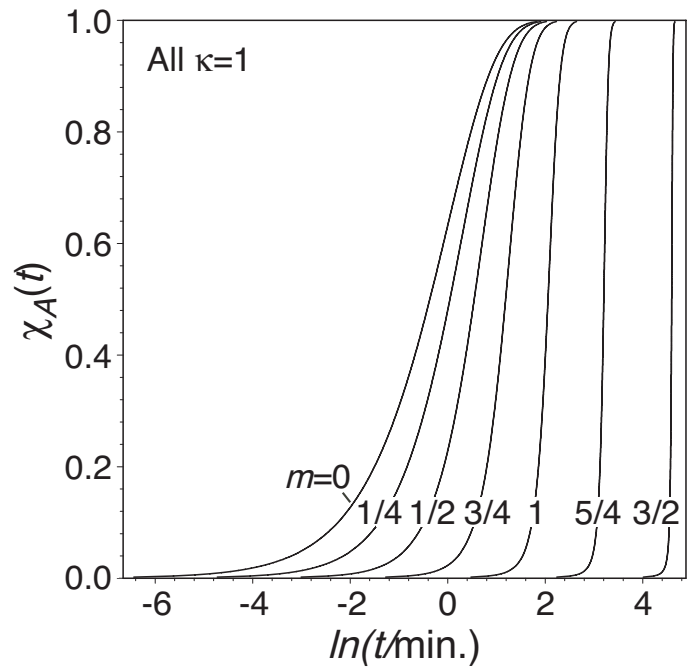

FIGURE 3a

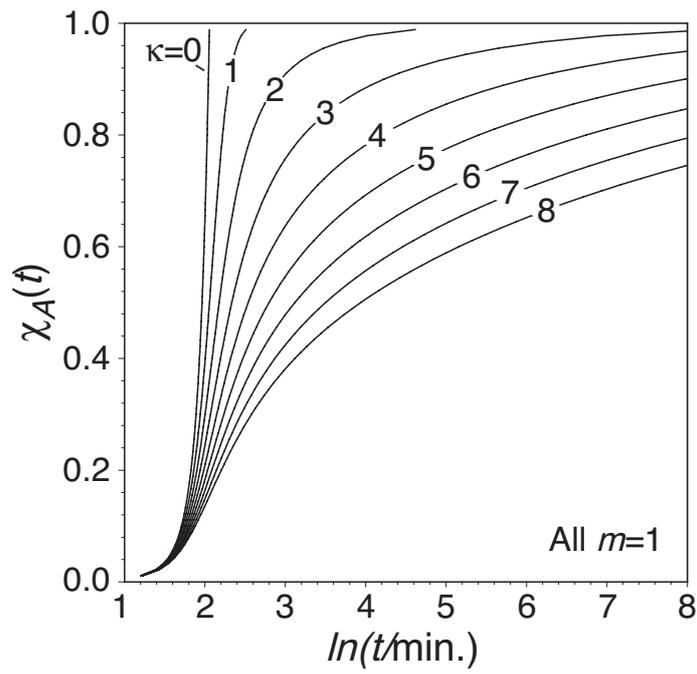

FIGURE 3b